\documentclass[twocolumn,aps,showpacs,prb,tightenlines,amsmath,amssymb]{revtex4}
\usepackage{bm}
\usepackage{graphicx}             
\usepackage{amssymb}               
\usepackage{amsmath}
\usepackage{colordvi}
\usepackage{calrsfs} 
\newcommand{\bgreek}[1]{\mbox{\boldmath$#1$\unboldmath}}
\begin{document}   

\title{Electron spin relaxation in graphene with random Rashba
  field: Comparison of D'yakonov-Perel' and Elliott-Yafet--like mechanisms}
 
\author{P. Zhang}
\author{M. W. Wu}
\thanks{Author to whom correspondence should be addressed}
\email{mwwu@ustc.edu.cn.}
\affiliation{Hefei National Laboratory for Physical Sciences at
  Microscale and Department of Physics, 
University of Science and Technology of China, Hefei,
  Anhui, 230026, China} 
\date{\today}

\begin{abstract} 
Aiming to understand the main spin
 relaxation mechanism in graphene, we investigate the spin relaxation 
with random Rashba field induced by both adatoms and substrate, by means of the
kinetic spin Bloch equation approach. 
 The charged adatoms on one hand
enhance the Rashba spin-orbit coupling locally and on the other hand serve as
Coulomb potential scatterers. Both effects contribute to 
spin relaxation limited by the D'yakonov-Perel' mechanism. In addition,
the random Rashba field also causes spin relaxation by spin-flip
scattering, manifesting itself as an Elliott-Yafet--like mechanism. 
Both mechanisms are sensitive to the correlation 
length of the random Rashba field,
 which may be affected by the environmental parameters such as
electron density and temperature.
By fitting and comparing the experiments from the
 Groningen group [J\'ozsa {\it et al.}, Phys. Rev. B {\bf 80},
 241403(R) (2009)] and Riverside group [Pi {\it et al.},
 Phys. Rev. Lett. {\bf 104}, 187201 (2010); Han and Kawakami, {\it ibid.} {\bf 107}, 047207 (2011)] which show 
either D'yakonov-Perel'-- (with the spin relaxation rate being inversely
proportional to the momentum scattering rate) or Elliott-Yafet--like
(with the spin relaxation rate being proportional to the momentum scattering 
rate) properties, we
suggest that the D'yakonov-Perel' mechanism dominates the spin relaxation in
graphene. The latest experimental finding of a
 nonmonotonic dependence of spin
  relaxation time on diffusion coefficient by Jo {\it et al.} 
 [Phys. Rev. B {\bf 84}, 075453 (2011)] is also well reproduced by our model.

\end{abstract}
\pacs{72.25.Rb, 71.70.Ej, 67.30.hj, 05.40.$-$a}
\maketitle

\section{Introduction}
Graphene, a two-dimensional allotrope of carbon with a honeycomb
lattice, has attracted much attention due to its two
dimensionality, Dirac-like energy spectrum and potential for the
all-carbon based electronics and spintronics in recent
years.\cite{novoselov666,Tombros_07,Geim_07,castro,kuemmeth,peres2673,mucciolo273201,sarma407} With the breaking of
inversion symmetry, possibly caused by ripples,\cite{Hernando_06} perpendicular
electric fields,\cite{Kane_SOC,Hernando_06,Min,Fabian_SOC,abde}
adsorbed adatoms,\cite{Castro_imp,abde,varykhalov} the 
substrate,\cite{ryu4944,Fabian_SR,dedkov107602} etc., the Rashba
spin-orbit coupling\cite{Rashba,Kane_SOC,Min} arises and results
 in spin relaxation
in the presence of scattering in graphene. A number of
 experiments on spin relaxation in graphene on SiO$_2$
substrate are available, revealing a spin relaxation time of the order
of 10-100~ps.\cite{Tombros_07,han222109,Pi,Jozsa_09,Tombros_08,Popinciuc,han1012.3435,yang047206,jo}
However, somewhat contradictory results are
exhibited in these experiments. While a decrease of spin relaxation rate with increasing  
momentum scattering rate has been observed by Riverside group
by surface chemical doping at 18~K,\cite{Pi} a linear scaling between the
momentum and spin scattering has been observed by both
Groningen group\cite{Jozsa_09,Popinciuc,Tombros_08} at room temperature and very recently 
Riverside group\cite{han1012.3435} at low temperature ($\le 100$~K) via varying the
electron density in graphene. The
D'yakonov-Perel'\cite{dp} (DP) mechanism was justified to be
important by the former phenomenon, however, the Elliott-Yafet\cite{ey} (EY)
mechanism was suggested to account for the latter. 
In addition, at a temperature as low as 4.2~K, a nonmonotonic dependence of spin
relaxation time on diffusion coefficient with the increase of
electron density was also reported by Jo {\it et al.} very
recently.\cite{jo} They claimed that the spin relaxation is due to the
EY mechanism. To fully understand
the spin relaxation in graphene, theoretical studies are in 
progress.\cite{hernando146801,Castro_imp,yzhou,Fabian_SR,dugaev085306,arxiv1107.3386,zhanggraphene}

Theoretically, the EY mechanism is revealed to be both invalid
  to account for the linear scaling between the momentum and spin
  scattering with the increase of electron density\cite{arxiv1107.3386} and also less
  important than the DP one.\cite{hernando146801} According to Refs.~\onlinecite{hernando146801} and
\onlinecite{arxiv1107.3386}, the spin relaxation times caused by the EY and
DP mechanisms are $\tau_s^{\rm EY}\approx (\hbar
v_fk_f/\lambda)^2\tau_p$ and $\tau_s^{\rm DP}\approx
\hbar^2/(\lambda^2\tau_p)$ respectively. Here $v_f$ and $k_f$ are the Fermi
velocity and momentum, $\lambda$ is the Rashba spin-orbit coupling
strength and $\tau_p$ is the momentum relaxation time. Consequently, $\tau_s^{\rm   EY}/\tau_p\propto
 k_f^2$,\cite{arxiv1107.3386,hernando146801} disagreeing with the
 electron density-independent linear scaling between $\tau_s$ and
 $\tau_p$ in the experiments.\cite{Jozsa_09,Popinciuc,Tombros_08,han1012.3435}
 The latest experiment by Jo {\it et al.} reveals the relation $\tau_s^{\rm   EY}/\tau_p\propto
 k_f^2$ with the increase of carrier density and hence suggests that the EY
 mechanism dominates spin relaxation in graphene.\cite{jo}
 Nonetheless, from the above theoretical results it is found
 that $\tau_s^{\rm EY}/\tau_s^{\rm DP}\approx (k_fl)^2$ with $l$ being the
 mean free path. Typically $(k_fl)^2\gg 1$ in graphene, meaning that the DP mechanism dominates
 over the  EY one.\cite{hernando146801} Due to
 the above reasons, the EY mechanism can not dominate spin relaxation in
 graphene.

The DP mechanism then becomes the reasonable candidate for the dominant
spin relaxation mechanism. Initially, the spin
relaxation time limited by the DP mechanism was calculated to be much longer than the
experimental data $\sim$10-100~ps due to the weak Rashba field. For example, ripples with curvature radii $\sim$
50~nm induce the local Rashba spin-orbit coupling with 
$\lambda\sim 8.5$~$\mu$eV,\cite{Hernando_06} and a perpendicular
electric field of magnitude $E_z$ ($E_z\sim 0.1$~V/nm) contributes a Rashba spin-orbit
coupling with $\lambda\sim\zeta E_z$, where $\zeta$ is
0.258~$\mu$eV/(V/nm) from rough estimation,\cite{Kane_SOC} 
17.9 or 66.6~$\mu$eV/(V/nm) from the tight-binding
model\cite{Hernando_06,Min} and 5~$\mu$eV/(V/nm) from the first-principles calculation.\cite{Fabian_SOC,abde} 
Based on this weak Rashba spin-orbit coupling, Zhou and Wu calculated spin
 relaxation in graphene on SiO$_2$ substrate with mobility $\sim
  10^4$~cm$^2$/(V$\cdot$s) taken from  the charge transport
 measurement\cite{novoselov666} and obtained a quite long spin
 relaxation time of the order of $\mu$s.\cite{yzhou} However, for 
the nonlocal  measurements of the spin relaxation,  
 the mobility is  at least  one order of magnitude 
smaller,\cite{Tombros_08,han222109,Popinciuc,Jozsa_09,han1012.3435,jo}
most likely caused by the extrinsic factors
induced by the  ferromagnetic electrodes,\cite{private}
 e.g., the adatoms.  The adatoms, as well as the influence of the substrate, 
may substantially enhance the Rashba spin-orbit 
coupling by distorting the graphene lattice and inducing $sp^3$ hybridization, leading $\lambda$ to be $\sim$ 
meV.\cite{Castro_imp,abde,varykhalov,dedkov107602} With this enhanced
Rashba spin-orbit coupling, the spin relaxation time in 
graphene is estimated\cite{Fabian_SR} and calculated\cite{zhanggraphene} to be comparable to the
experimental data. Nevertheless, whether and how the DP mechanism accounts
for the experimentally observed linear scaling between the momentum
and spin scattering is questionable. 

Apart from the above two mechanisms, another EY-like mechanism may
contribute to the spin relaxation in graphene when the Rashba field
is random in the real space. As proposed by Sherman in
semiconductors,\cite{sherman209,sherman67,glazov2157} 
the randomness of spin-orbit coupling
contributes to or even dominates spin relaxation by spin-flip scattering under certain
conditions.\cite{sherman209,sherman67,glazov2157,zhou57001,dugaev085306}
For graphene, the Rashba field induced by a fluctuating electric 
field from ionized impurities in the substrate or ripples is indeed
random in the real space. The former case, with the average  
Rashba field being nonzero, has been investigated by Ertler {\it
  et al.} via Monte Carlo simulation,\cite{Fabian_SR} while the latter
one, with the average Rashba field being zero, has been studied by
Dugaev {\it et al.}\cite{dugaev085306} via the kinetic
equations. However, for both cases the calculated spin
relaxation time is much longer than the experimental
data.

In this work, we investigate spin relaxation in graphene with random
Rashba field (RRF) induced by adatoms and substrate by means of the
kinetic spin Bloch equation (KSBE) approach.\cite{wu-review} A random Rashba model is set up, where the charged
adatoms on one hand enhance the Rashba spin-orbit
coupling locally and on the other hand serve as Coulomb potential
scatterers. Based on this model, an analytical study on spin relaxation with RRF 
is performed. It is found that while the average Rashba field leads to
spin relaxation limited by the DP mechanism, which is absent in the
work by Dugaev {\it et al.},\cite{dugaev085306} the randomness causes spin
relaxation via spin-flip scattering. With the increase of adatom
density, the spin relaxation caused by the spin-flip scattering due to
the RRF always shows an EY-like behaviour 
(the spin relaxation rate is proportional to the 
momentum scattering rate) whereas the DP mechanism can
 exhibit either EY- or DP-like
(the spin relaxation rate is inversely proportional to the 
momentum scattering rate) one. When all the other parameters are
fixed, with the increase of
electron density the spin
relaxation rates due to both mechanisms increase;
Nevertheless, the spin relaxation rate 
determined by the spin-flip scattering due to the RRF
is insensitive to the temperature whereas that determined by the DP
mechanism becomes insensitive to the temperature when the electron-impurity
scattering is dominant. However, the correlation length of the RRF may
vary with the electron density as well as temperature and both mechanisms are
 sensitive to the correlation length.

We carry out numerical calculations and fit the experiments of
Riverside\cite{Pi,han1012.3435} and Groningen\cite{Jozsa_09}
groups. By fitting the DP-like behaviour with the increase of adatom
density observed by the Riverside group,\cite{Pi} we find that only when
the DP mechanism is dominant can the experimental data be 
understood and the effect of the spin-flip scattering due to the RRF is
negligible. However, the experimental EY-like
behaviour with the increase of electron density first observed
by the Groningen group\cite{Jozsa_09} can be fitted from  our model
with either the DP mechanism or the spin-flip scattering
due to the RRF being dominant by taking into account the decrease of the 
correlation length of the RRF with the
increase of electron density. Nevertheless, the fact
that the Riverside group has also observed the similar EY-like behaviour
in their samples very recently\cite{han1012.3435} suggests that the DP mechanism
is dominant but exhibits EY-like properties with the increase of
electron density. The temperature dependence of the spin relaxation from
the Riverside group,\cite{han1012.3435} suggested to be the evidence of
the EY mechanism, is also fitted by our random Rashba model with the
DP mechanism being dominant. The corresponding temperature dependence from the 
spin-flip scattering due to the RRF is  demonstrated to be in fact
temperature insensitive. The similar experimental phenomenon observed with the variation of the electron
  density by the two groups further
suggests that the DP mechanism also dominates the 
spin relaxation in the experiment of Groningen group.\cite{Jozsa_09}
Moreover, the latest reported nonmonotonic dependence of spin relaxation time on
  diffusion coefficient\cite{jo} is also fitted.

This paper is organized as follows. In Sec.~II, we present the model
and introduce the KSBEs. In Sec.~III, we investigate the spin
relaxation analytically and discuss the relative importance of the DP
mechanism and mechanism of the spin-flip scattering due to the RRF. 
In Sec.~IV, we perform numerical
calculations and fit the experimental data. We discuss and 
summarize in Sec.~V.

\section{Model and KSBEs}
The $n$-doped graphene monolayer under investigation lies on the SiO$_2$
substrate perpendicular to the $z$-axis. 
The random Rashba spin-orbit coupling reads\cite{Kane_SOC,Min}
\begin{equation}
H_{\rm so}=\lambda({\bf r})(\mu\tau_x\sigma_y-\tau_y\sigma_x).
\end{equation}
Here $\mu=\pm 1$ labels the valley located at $K$ or $K^\prime$. 
${\bgreek \tau}$ and ${\bgreek \sigma}$ are the Pauli matrices in
the sublattice and spin spaces, respectively. The position-dependent coupling strength $\lambda({\bf
  r})$, mainly contributed by the randomly
distributed adatoms and also possibly by the substrate, can be modeled as
\begin{equation}
  \lambda({\bf r})=\lambda_0^i+\sum_{n=1}^{N^{a}_i}\delta_n e^{-|{\bf r}-{\bf
        R}_n|^2/2\xi^2}.
    \label{lambdar}
  \end{equation}
Here the second term is contributed by the adatoms with a total number
$N_i^a$. In this model it is assumed that an adatom located at ${\bf
  R}_n$ induces a local Rashba field peaking at ${\bf R}_n$ 
with a magnitude of $\delta_n$ and decaying within a length scale $\sim \xi$ following the Gaussian
form. $\delta_n$ is of the order of meV (Refs.~\onlinecite{Castro_imp,abde,varykhalov}) and $\xi$ is larger than the
graphene lattice constant 0.25~nm.\cite{Castro_imp} The first term $\lambda_0^i$ comes from the average
contribution from the substrate, whose fluctuation is phenomenally
incorporated by affecting $\xi$. The mean value of $\lambda({\bf r})$ reads
$\lambda_0\equiv\langle\lambda({\bf r})\rangle=\lambda_0^i+\lambda_0^a$ with
$\lambda_0^a=2\pi n_i^a\overline{\delta}\xi^2$, where $n_i^a$ is the areal density of adatoms. The correlation
function $C({\bf r})\equiv\langle[\lambda({\bf
  r})-\lambda_0][\lambda(0)-\lambda_0]\rangle=\pi
  n_i^a\overline{\delta^2}\xi^2e^{-r^2/4\xi^2}$, with the corresponding Fourier
  transformation 
\begin{equation}
C_{\bf q}=4\pi^2n_i^a\overline{\delta^2}\xi^4e^{-\xi^2q^2}.
\label{cqeq}
\end{equation}
This expression is similar to that given in
  Ref.~\onlinecite{dugaev085306} except that it depends on
  $\xi$ in a higher order here. For single-sided
  adatoms (the adatoms are distributed on the graphene surface), we choose $\delta_n=\delta$, by which
  $\overline{\delta}=\delta$ and $\overline{\delta^2}=\delta^2$. For double-sided
  adatoms (the adatoms are distributed both on the graphene surface and
  at the graphene/substrate interface), we set $\delta_n=\delta_1$ or $\delta_2$
  ($\delta_1\delta_2<0$) with equal possibilities, by which
  $\overline{\delta}=0.5(\delta_1+\delta_2)$ and
  $\overline{\delta^2}=0.5(\delta_1^2+\delta_2^2)$. It is noted that
  the random Rashba model proposed here is modified from the short-range random
  potential model depicting the electron-hole puddles in
  graphene.\cite{lewen081410}

Under the basis laid out in Refs.~\onlinecite{Fabian_SR} and 
\onlinecite{yzhou}, the electron Hamiltonian can be written as\cite{yzhou} 
\begin{eqnarray}
    H = \sum_{{\mu}{\bf k}ss^\prime}
  \big[\varepsilon_{\bf k} \delta_{ss^\prime}
  +\bm{\Omega}_{\bf k}\cdot{\bgreek\sigma}_{ss^\prime}\big] {c_{\mu{\bf k}s}}^{\dagger}
  c_{\mu{\bf k}s^\prime} + {H}_{\rm int}
  \label{H_eff}
\end{eqnarray}
in the momentum space. Here $c_{\mu{\bf k}s}$ (${c_{\mu{\bf k}s}}^{\dagger}$) is
the annihilation (creation) operator of electrons in the $\mu$ valley
with momentum ${\bf k}$ (relative to the valley center) and spin $s$
($s=\pm \frac{1}{2}$). $\varepsilon_{\bf k}=\hbar v_fk$ with
$v_f=10^6$~m/s. The effective magnetic field $\bm{\Omega}_{\bf
  k}$ from the average Rashba field is
\begin{eqnarray}
\bm{\Omega}_{\bf k}=\lambda_0(-\sin\theta_{\bf k}, \cos\theta_{\bf
  k},0),
\label{omega}
\end{eqnarray}
where $\theta_{\bf k}$ is the polar
angle of momentum ${\bf k}$. The Hamiltonian $H_{\rm int}$ consists of the spin-conserving scattering
[electron-impurity\cite{Adam_08} (here the impurities include both the ones existing
in the substrate and the charged adatoms,\cite{Pi,mccreary} taken into
account by the minimal model proposed by Adam and Das Sarma\cite{Adam_08}),
electron-phonon,\cite{hwang1,lazzeri,fratini} and electron-electron\cite{yzhou} scattering] as well
as the spin-flip scattering due to the RRF,\cite{dugaev085306,glazov2157}
\begin{eqnarray}
H_{\rm flip}=\sum_{\mu,{\bf k^\prime}\neq{\bf k},ss^\prime}\lambda_{{\bf
    k}-{\bf k^\prime}}V_{{\bf kk^\prime}ss^\prime}c_{\mu{\bf k}s}^\dagger
c_{\mu{\bf k}^\prime s^\prime},
\end{eqnarray}
where
\begin{eqnarray}
\lambda_{\bf q}=\int \lambda({\bf r})e^{-i{\bf  q}\cdot{\bf r}}d{\bf r}
\end{eqnarray}
and
\begin{eqnarray}
V_{\bf kk^\prime}=\left(\begin{array}{cc}
0 & -ie^{-i\theta_{\bf k}}\\
ie^{i\theta_{\bf k^\prime}} & 0
\end{array}\right). 
\end{eqnarray}

The KSBEs are\cite{wu-review}
\begin{equation}
  \partial_t \rho_{\mu\bf k}(t)=\partial_t\rho_{\mu\bf k}(t)|_{\rm coh}+\partial_t\rho_{\mu\bf k}(t)|_{\rm  scat}^c+\partial_t\rho_{\mu\bf k}(t)|_{\rm  scat}^f.
\label{ksbee}
\end{equation}
Here $\rho_{\mu\bf k}(t)$ represent the density matrices of
electrons with relative momentum ${\bf k}$ in valley ${\mu}$ at
 time $t$. The coherent terms $\partial_t\rho_{\mu\bf k}(t)|_{\rm
   coh}=-\frac{i}{\hbar}[\bm{\Omega}_{\bf
   k}\cdot{\bgreek\sigma},\rho_{\mu   {\bf k}}(t)]$ with the Hartree-Fock term from the Coulomb
interaction being neglected due to the small spin
polarization.\cite{wu-review,yzhou} The concrete expressions of
the spin-conserving scattering terms $\partial_t\rho_{\mu\bf k}(t)|_{\rm  scat}^c$ can be found in
Ref.~\onlinecite{yzhou}. When the electron mean free path $l$ is much
longer than the correlation length of the fluctuating Rashba field
$\xi$ (this is easily satisfied as $\xi\sim$ nm while
$l\sim$ 10-100~nm in graphene), i.e., the electron spins experience
indeed the random spin-orbit coupling, the spin-flip scattering terms can
be written as (Appendix~\ref{app-a})\cite{dugaev085306,glazov2157}

\begin{eqnarray}\nonumber
\partial_t\rho_{\mu{\bf k}}(t)|_{\rm scat}^{f}&=&-\frac{2\pi}{\hbar}\sum_{{\bf k}^\prime} C_{\bf
  k-k^\prime}\delta(\varepsilon_{\mu\bf k}-\varepsilon_{\mu\bf 
  k^\prime})\\  &&\mbox{}\times [\rho_{ \mu{\bf k}}(t)-V_{\bf kk^\prime}\rho_{\mu{\bf
  k^\prime}}(t)V_{\bf k^\prime k}].
\label{flip}
\end{eqnarray}
By solving the KSBEs, one can obtain the spin relaxation
properties from the time evolution of $\rho_{\mu{\bf
    k}}(t)$. 

\section{Analytical study of spin relaxation}
\label{analy}
In this section we analytically study the spin relaxation in graphene
with the RRF. To realize this, we only take into
account the spin-flip scattering as well as the strong
elastic electron-impurity scattering. When the valley index is further
omitted due to the degeneracy, the KSBEs are simplified to be 
\begin{eqnarray}\nonumber
\partial_t\rho_{{\bf k}}(t)=&&-\frac{i}{\hbar}[\bm{\Omega}_{\bf  k}\cdot{\bgreek\sigma},\rho_{
  {\bf k}}(t)]-\frac{2\pi}{\hbar}\sum_{{\bf k}^\prime} |U_{\bf
  k-k^\prime}|^2I_{\bf kk^\prime}\\ \nonumber
&&\mbox{}\times\delta(\varepsilon_{\bf
  k}-\varepsilon_{\bf k^\prime})[\rho_{{\bf k}}(t)-\rho_{{\bf
    k^\prime}}(t)]\\ \nonumber&&\mbox{}-\frac{2\pi}{\hbar}\sum_{{\bf
    k}^\prime} C_{{\bf k}-{\bf k^\prime}}\delta(\varepsilon_{\bf k}-\varepsilon_{\bf
  k^\prime})\\  &&\mbox{}\times [\rho_{ {\bf k}}(t)-V_{\bf kk^\prime}\rho_{{\bf
  k^\prime}}(t)V_{\bf k^\prime k}]
\label{sksbe}
\end{eqnarray}
Here $|U_{\bf k-k^\prime}|^2$ is the effective electron-impurity scattering
matrix element and $I_{\bf kk^\prime}=\frac{1}{2}[1+\cos(\theta_{\bf
    k}-\theta_{\bf k^\prime})]$ is the form factor.\cite{yzhou} $|U_{\bf
    k-k^\prime}|^2=n_i^a|U_{\bf k-k^\prime}^a|^2+n_i^s|U_{\bf
    k-k^\prime}^s|^2$, with the first (second) term corresponding to
  the scattering of electrons from adatoms (impurities in the substrate). $|U_{\bf
    k-k^\prime}^{a/s}|^2$ is the electron-impurity Coulomb potential scattering
  matrix element.\cite{Adam_08,yzhou,zhanggraphene} $n_i^s$ is the
  impurity density in the substrate. By defining
  the spin vector as ${\bf S}_{{\bf k}}(t)=\mbox{Tr}[\rho_{{\bf
      k}}(t){\bgreek \sigma}]$, one obtains the equation of
 ${\bf S}_{{\bf k}}(t)$ directly from
Eq.~(\ref{sksbe}) as 
\begin{eqnarray}\nonumber
\partial_t{\bf S}_{{\bf
    k}}(t)=&&\frac{2\lambda_0}{\hbar}{\bf A}_{\bf k}\cdot{\bf
  S}_{\bf k}(t)-\frac{2\pi}{\hbar}\sum_{{\bf k}^\prime} |U_{\bf
  k-k^\prime}|^2I_{\bf kk^\prime}\\ \nonumber
&&\mbox{}\times\delta(\varepsilon_{\bf
  k}-\varepsilon_{\bf k^\prime})[{\bf S}_{{\bf k}}(t)-{\bf S}_{{\bf
    k^\prime}}(t)]\\ \nonumber&&\mbox{}-\frac{2\pi}{\hbar}\sum_{{\bf k}^\prime} C_{{\bf
  k}-{\bf k^\prime}}\delta(\varepsilon_{\bf k}-\varepsilon_{\bf
  k^\prime})\\  &&\mbox{}\times [{\bf S}_{\bf k}(t)-{\bf
  B}_{\bf kk^\prime}\cdot{\bf  S}_{\bf k^\prime}(t)].
\label{analy-ksbe}
\end{eqnarray}
Here
\begin{eqnarray}
{\bf A}_{\bf k}=\left(
  \begin{array}{ccc}
    0 & 0 & \cos\theta_{\bf k}\\
    0 & 0 & \sin\theta_{\bf k}\\
    -\cos\theta_{\bf k} & -\sin\theta_{\bf k} & 0
    \end{array}
\right)
\end{eqnarray}
and 
\begin{eqnarray}
{\bf B}_{\bf kk^\prime}=\left(
  \begin{array}{ccc}
    -\cos(\theta_{\bf k}+\theta_{\bf k^\prime}) &  -\sin(\theta_{\bf
      k}+\theta_{\bf k^\prime}) & 0\\
    -\sin(\theta_{\bf k}+\theta_{\bf k^\prime}) &  \cos(\theta_{\bf k}+\theta_{\bf k^\prime}) & 0\\
    0 & 0 & -1
    \end{array}
\right).
\end{eqnarray}
By expanding ${\bf S}_{\bf k}(t)$ as ${\bf S}_{\bf  k}(t)=\sum_{l}{\bf
  S}_{k}^l(t)e^{il\theta_{\bf k}}$ and retaining the lowest three
orders of ${\bf S}_{k}^l(t)$ (i.e., terms with $l=0$, $\pm 1$), one obtains a group of differential
equations of ${\bf S}_{k}^{\pm 1,0}(t)$ (Appendix~\ref{app-b}). With
initial conditions, these equations can be solved and the information
on spin relaxation is obtained from $S_k^0(t)$. We label the spin relaxation rate
along the $x$-, $y$-, or $z$-axis for states with momentum $k$ as $\Gamma_x$,
$\Gamma_y$ or $\Gamma_z$, respectively. One has (Appendix~\ref{app-b})
\begin{equation}
\Gamma_z=2\Gamma_x=2\Gamma_y=2/\tau_{ks}^0+4\lambda_0^2\tau_k^1/\hbar^2,
\label{sr-rate}
\end{equation}
where
\begin{equation}
\frac{1}{\tau^1_k}=\frac{k}{4\pi\hbar^2v_f}\int_0^{2\pi}d\theta
|U_{\bf q}|^2\sin^2\theta
\label{taukl1}
\end{equation}
and 
\begin{equation}
\frac{1}{\tau^0_{ks}}=\frac{k}{2\pi\hbar^2v_f}\int_0^{2\pi}d\theta C_{\bf q},
\label{tauksl0}
\end{equation}
with $|U_{\bf q}|^2$ and $C_{\bf q}$ depending only on $|{\bf
  q}|=2k\sin\frac{\theta}{2}$. For a highly degenerate electron system
in graphene, the spin relaxation is contributed by the spin-polarized
electrons around the Fermi circle. Therefore, one can approximately obtain the spin
relaxation rate of the whole electron system by replacing $k$ with
$k_f$ in Eq.~(\ref{sr-rate}). 

From Eq.~(\ref{sr-rate}), one notices that the spin relaxation rate
(take $\Gamma_z$ as an example)
consists of two parts, $\Gamma_{\rm flip}=2/\tau_{k_fs}^0$, determined
by the spin-flip scattering due to the RRF, and
$\Gamma_{\rm DP}=4\lambda_0^2\tau_k^1/\hbar^2$, determined by the
average Rashba field due to the DP mechanism. These two
mechanisms contribute to spin relaxation independently. It is noted that
$\Gamma_{\rm flip}$ obtained here is consistent with that given in
Ref.~\onlinecite{dugaev085306} except the correlation functions are
different and $\Gamma_{\rm DP}$ is the one previously given in
Ref.~\onlinecite{zhanggraphene}. In the following we  discuss the spin
relaxation due to the two mechanisms respectively and compare their relative importance. 

\subsection{Spin relaxation caused by the spin-flip scattering due to
  the RRF}
\label{shermansection}
By utilizing Eq.~(\ref{cqeq}), one can obtain
the spin relaxation rate along the $z$-axis due to the spin-flip
scattering as
\begin{eqnarray}\nonumber
  \Gamma_{\rm flip}&=&\frac{k_f}{\pi\hbar^2v_f}\int_0^{2\pi}d\theta
  C_{\bf q}\\\nonumber &=&\frac{8\pi^2n_{i}^a\overline{\delta^2}\xi^4k_f}{\hbar^2 v_f}e^{-2\xi^2k_f^2}I_0(2\xi^2k_f^2)\\
  &=&\frac{8\pi^2n_{i}^a\overline{\delta^2}\xi^3}{\hbar^2 v_f}F(\xi k_f)
  =\frac{8\pi^2n_{i}^a\overline{\delta^2}k_f^{-3}}{\hbar^2 v_f}G(\xi k_f).
\label{fxgx}
\end{eqnarray}
In the limits $k_f\xi\ll 1$ and $k_f\xi\gg 1$,
 \begin{eqnarray}
    \Gamma_{\rm flip}\approx\frac{8\pi^2n_{i}^a\overline{\delta^2}}{\hbar^2 v_f}
\begin{cases}
\xi^4k_f & (k_f\xi\ll 1) \\
\xi^3/(2\sqrt{\pi})  &(k_f\xi\gg 1)
\end{cases}.
\label{flip-limit}
\end{eqnarray}
In the above equations $I_0(x)$ is the modified Bessel
function, $F(x)=xe^{-2x^2}I_0(2x^2)$ and
$G(x)=x^3F(x)$. 

\begin{figure}[ht]
  {\includegraphics[width=8cm]{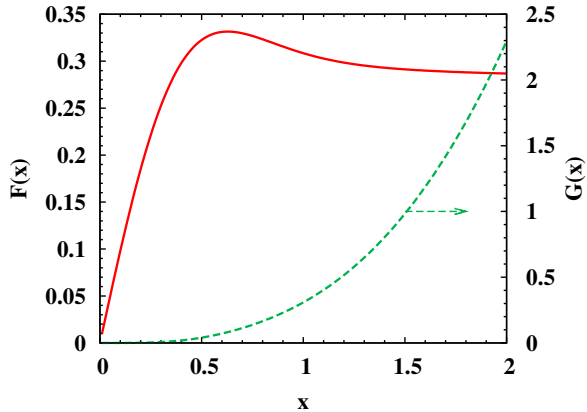}}
  \caption{(Color online) $F(x)$ and
    $G(x)$ in Eq.~(\ref{fxgx}). The scale of $G(x)$ is on the right-hand side of the frame.}
  \label{figzw1}
\end{figure}

We now focus on the various factors affecting $\Gamma_{\rm
flip}$.  $\Gamma_{\rm flip}$ is proportional to the adatom density $n_i^a$, as
expected. From the $x$ dependence of $F(x)$
and $G(x)$ as shown in Fig.~\ref{figzw1},
 one can explore the dependence of $\Gamma_{\rm flip}$ on
$k_f$ and $\xi$, respectively. It is found that with the increase of $k_f$,
$\Gamma_{\rm flip}$ first increases almost linearly when $k_f\xi\le
0.63$, then decreases mildly and eventually saturates
[$F(+\infty)=\frac{1}{2\sqrt{\pi}}$]. As a result, with the increase
of electron density $n_e$, $\Gamma_{\rm flip}$ has a nonmonotonic
behaviour with a peak located at
$n_e^c=\frac{1}{\pi}(\frac{0.63}{\xi})^2$. When $n_e\ll n_e^c$,
$\Gamma_{\rm flip}$ is proportional to  $n_e^{1/2}$; and when $n_e\gg
n_e^c$, $\Gamma_{\rm flip}$ becomes insensitive to $n_e$. These
features are in consistence with those presented in
Ref.~\onlinecite{dugaev085306}. We give an estimation on
$n_e^c$ here. When the correlation length $\xi$ is set as 1~nm, $n_e^c\approx 1.3\times
10^{13}$~cm$^{-2}$, which is a quite high value compared to the
experimental data. Usually $n_e$ varies around
10$^{12}$~cm$^{-2}$. Therefore, in order to observe the nonmonotonic
behaviour of $\Gamma_{\rm flip}$ with increasing $n_e$, $\xi$ is
required to be around a relatively large value, e.g.,
3.5~nm. When the electron density is fixed, $\Gamma_{\rm flip}$
increases monotonically with increasing $\xi$, as indicated by the $x$
dependence of $G(x)$ in Fig.~\ref{figzw1}. The effect of temperature $T$
on $\Gamma_{\rm flip}$ can be inferred from the $k$ dependence of
$\Gamma_{\rm flip}$. For the highly degenerate electron system,
$\Gamma_{\rm flip}$ is insensitive to $T$. When the electron density
is relatively low, with increasing $T$, $\Gamma_{\rm flip}$ is expected to increase mildly as
electrons tend to occupy the states with larger momentum. With typical values
$\overline{\delta^2}=25$~meV$^2$, $n_i^a=3\times 10^{12}$~cm$^{-2}$,
$n_e=10^{12}$~cm$^{-2}$ and $\xi=0.5$~nm, one has $k_f\xi\approx
0.09$ and $\Gamma_{\rm flip}^{-1}\approx 670$~ps. If $\xi$ is changed
to be 2 times larger, i.e., 1~nm, $\Gamma_{\rm flip}^{-1}$ becomes about 16 times smaller, as
$\Gamma_{\rm flip}$ is proportional to $\xi^4$ when $k_f\xi\ll 1$ [Eq.~(\ref{flip-limit})]. 

\subsection{Spin relaxation caused by the DP mechanism}
\label{dpsection}
In the analytical study, only the elastic spin-conserving scattering
is considered. With the other spin-conserving scattering included, the spin
relaxation rate due to the DP mechanism should be modified to be 
\begin{equation}
\Gamma_{\rm DP}=4\lambda_0^2\tau_p^\ast(k_f)/\hbar^2, 
\end{equation}
where $\tau_p^\ast(k_f)$ is the effective momentum relaxation time limited by all
the different kinds of scattering, including the electron-electron
Coulomb scattering.\cite{wu-review} When $\lambda_0$ is fixed, the
electron density and temperature affect $\Gamma_{\rm DP}$ via $\tau_p^\ast(k_f)$. 
It has been shown previously\cite{yzhou} that with the increase of $n_e$ or the decrease of $T$, the
scattering strength decreases and $\Gamma_{\rm DP}$
increases. However, when the electron-impurity scattering
is dominant, $\Gamma_{\rm DP}$ is
insensitive to $T$. The dependence of $\Gamma_{\rm DP}$ on $n_i^a$ 
is not obvious. To facilitate the investigation, we write ${\tau_p^\ast}^{-1}(k_f)=\tau_{p,i}^{-1}(k_f)+\tau_{p,a}^{-1}(k_f)$, where
$\tau_{p,a}^{-1}(k_f)$ is contributed by the electron-charged adatom
scattering and $\tau_{p,i}^{-1}(k_f)$ by all the other scattering. From Eq.~(\ref{taukl1}), one has 
\begin{equation}
\tau_{p,a}^{-1}(k_f)=\frac{n_i^ak_f}{4\pi\hbar^2v_f}\int_0^{2\pi}d\theta |U^a_{\bf q}|^2\sin^2\theta\equiv c_1n_i^a
\label{tauadatom} 
\end{equation}
with $|{\bf q}|=2k_f\sin\frac{\theta}{2}$. With $\lambda_0=\lambda_0^i+2\pi
n_i^a\overline{\delta}\xi^2\equiv\lambda_0^i+c_2n_i^a$, $\Gamma_{\rm
  DP}$ can be written as 
\begin{equation}
\Gamma_{\rm 
  DP}=\frac{4}{\hbar^2}\frac{(\lambda_0^i+c_2n_i^a)^2}{\tau_{p,i}^{-1}(k_f)+c_1n_i^a},
\label{gammadp}
\end{equation}
which indicates a complex dependence on $n_i^a$. When $c_2= 0$ and
$\lambda_0^i\neq 0$, $c_2>0$ and $0\le n_i^a\le
c_2^{-1}\lambda_0^i-2c_1^{-1}\tau_{p,i}^{-1}(k_f)$ or $c_2<0$ and
$0\le n_i^a\le -c_2^{-1}\lambda_0^i$, $\Gamma_{\rm DP}$ decreases with
increasing $n_i^a$, exhibiting the DP-like behaviour. However, most
interestingly, when  $c_2>0$ and $n_i^a\ge
c_2^{-1}\lambda_0^i-2c_1^{-1}\tau_{p,i}^{-1}(k_f)$ or $c_2<0$ and
$n_i^a\ge -c_2^{-1}\lambda_0^i$, $\Gamma_{\rm DP}$ increases with
increasing $n_i^a$, exhibiting the EY-like behaviour. Particularly,
here in the limit with $n_i^a$ being
large enough, $\Gamma_{\rm DP}\propto n_i^a$ approximately. With typical values in the
  presence of adatoms,\cite{zhanggraphene} $\lambda_0=0.2$~meV and
  $\tau_p^\ast(k_f)=0.1$~ps, $\Gamma_{\rm DP}^{-1}$ is calculated to be
  about 100~ps.

\subsection{Comparison of relaxations caused by the spin-flip scattering
  due to the RRF and the DP mechanism}
In this subsection we discuss the relative importance of the mechanism of spin-flip
scattering due to the RRF and the DP mechanism in the regime with $k_f\xi\ll 1$, which is typically realized in
graphene. Under this condition, from Eqs.~(\ref{flip-limit}) and (\ref{gammadp}), one obtains $\Gamma_{\rm DP}/\Gamma_{\rm
    flip}\approx 10{\overline{\delta}}^2/\overline{\delta^2}$ when
  $\lambda_i^0=\tau_{p,i}^{-1}(k_f)=0$. For the case with single-sided
  adatoms, ${\overline{\delta}}^2=\overline{\delta^2}$, therefore
  $\Gamma_{\rm DP}/\Gamma_{\rm flip}\approx 10$ and $\Gamma_{\rm flip}$ can be neglected. However, for the case with
  double-sided adatoms, $\Gamma_{\rm flip}$ can be comparable to or
  even surpass $\Gamma_{\rm DP}$ as $\overline{\delta}$ may be as
  small as zero. In reality $\tau_{p,i}^{-1}(k_f)\neq 0$ and
  $\Gamma_{\rm DP}$ decreases with increasing $\tau_{p,i}^{-1}(k_f)$. When
  the substrate also contributes to the Rashba field as $\lambda_i^0$, $\Gamma_{\rm
    DP}$ can be either enhanced (e.g., when
  $\lambda_i^0\overline{\delta}>0$) or suppressed. As a consequence,
  for the configuration with single-sided adatoms, when the 
  contribution from the substrate to the average Rashba field
  does not compensate that from the adatoms (e.g., when $\lambda_i^0\overline{\delta}\ge
  0$) and the scattering other than the 
  electron-adatom type is not extraordinarily strong (i.e.,
  $\tau_{p,i}^{-1}$ is not unusually large), the spin relaxation
  caused by the spin-flip scattering due to the RRF can
  be neglected. In such a case,  
  the spin relaxation is limited by the DP mechanism with the adatoms
  contributing to the average Rashba field. This is just how
  the effect of adatoms was incorporated in our previous investigation.~\cite{zhanggraphene} For other cases,
  whether the spin-flip scattering due to the RRF is important or not when compared to
  the DP mechanism is condition-dependent. Undoubtly, when the average Rashba field approaches
  zero, the spin-flip scattering due to the RRF tends to be important. In the 
  work of Dugaev {\it et al.},\cite{dugaev085306} the average Rashba
  field induced by ripples is zero and the spin relaxation is solely
  determined by the spin-flip scattering due to the RRF. However, the spin relaxation time calculated
  in their model is of the order of 10~ns, two orders of magnitude
  larger than the experimental values.

\section{Numerical results}
The KSBEs need to be solved numerically in order to take full account
of all the different kinds of scattering. The initial conditions are set as
\begin{eqnarray}
&&\rho_{\mu \bf k}(0)=\frac{f^0_{{\bf k}\uparrow}+f^0_{{\bf
      k}\downarrow}}{2}+\frac{f^0_{{\bf k}\uparrow}-f^0_{{\bf
      k}\downarrow}}{2}{\bf\hat n}\cdot{\bgreek \sigma},
\label{i1}\\
&&\sum_{\mu\bf  k}\mbox{Tr}[\rho_{\mu\bf k}(0){\bf
  \hat{n}}\cdot{\bgreek \sigma}]=n_eP(0),
\label{i2}\\
&& \sum_{\mu\bf
  k}\mbox{Tr}[\rho_{\mu\bf k}(0)]=n_e. 
\label{i3}
\end{eqnarray}
At time $t=0$, the electrons are polarized along ${\bf\hat n}$ with
the density and spin polarization being $n_e$ and $P(0)$,
respectively. $f^0_{{\bf k}\uparrow/\downarrow}$ is the Fermi
distribution function of electrons with spins parallel/antiparallel to
${\bf\hat n}$, where the
chemical potential is determined by
Eqs.~(\ref{i2})-(\ref{i3}). By solving the KSBEs, one can obtain
the time evolution of spin polarization along ${\bf\hat n}$ as
$P(t)=\frac{1}{n_e}\sum_{\mu\bf  k}\mbox{Tr}[\rho_{\mu\bf k}(t){\bf
  \hat{n}}\cdot{\bgreek \sigma}]$ and hence the spin relaxation
time $\tau_s$. In the calculation, we set $P(0)$ to be as small as 0.05 and ${\bf\hat n}$
in the graphene plane, such as, along the $y$-axis, in order to compare
with experiments. 

\subsection{Adatom density dependence of spin relaxation}
\begin{figure}[ht]
  {\includegraphics[width=8cm]{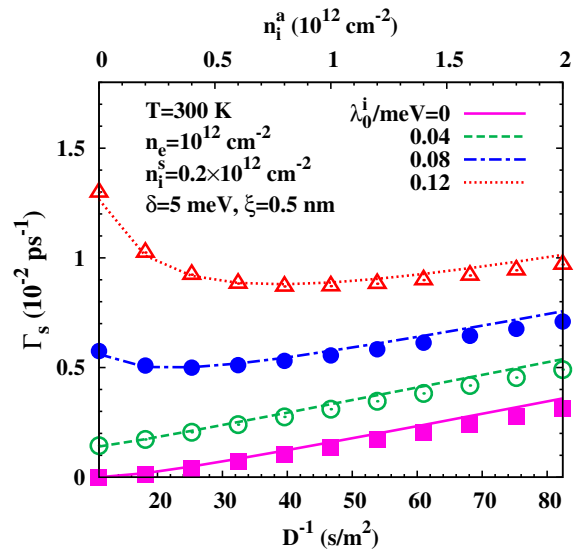}}
  \caption{(Color online) The dependence of in-plane spin relaxation
    rate  on the adatom density (at the top of the frame) or 
    the inverse of the charge diffusion coefficient (at the bottom of
    the frame). The temperature is $T=300$~K, the electron density is $n_e=10^{12}$~cm$^{-2}$,
    the density of impurities in the substrate is $n_i^s=0.2\times
    10^{12}$~cm$^{-2}$ and the parameters for the single-sided adatom model
    are $\delta=5$~meV and $\xi=0.5$~nm. The curves are calculated
    with different $\lambda_0^i$. For each curve, its nearby data
    points are calculated with the same parameters but without the 
    spin-flip scattering.} 
  \label{figzw2}
\end{figure}

In this section we study the adatom density dependence of spin
relaxation based on the single-sided adatom model. In
Fig.~\ref{figzw2}, the in-plane spin relaxation rate against the
adatom density (at the top of the frame) or the inverse of charge
diffusion coefficient (at the bottom of the frame) is shown. The
temperature is $T=300$~K, the electron density is $n_e=10^{12}$~cm$^{-2}$,
the density of impurities in the substrate is $n_i^s=0.2\times
10^{12}$~cm$^{-2}$ and the parameters for the single-sided adatom model
are $\delta=5$~meV and $\xi=0.5$~nm. In the figure, the spin
relaxation rates with different values of $\lambda_0^i$ are plotted by the
curves. The nearby data points of each curve are 
calculated with the spin-flip scattering being removed.
 The small discrepancy between each curve and corresponding
data points indicates that the DP mechanism dominates the spin
relaxation. It is noted that in a large parameter regime of the
background Rashba field $\lambda_i^0$, the curves show obvious EY-like behaviour, i.e., the
spin relaxation rate is proportional to the momentum relaxation
rate. When $\lambda_i^0$ is large enough
[larger than $2c_2c_1^{-1}\tau_{p,i}^{-1}(k_f)\approx 0.05$~meV from the discussion in
Sec.~\ref{dpsection}], the spin relaxation rate decreases with
increasing adatom density at low doping density of adatoms, showing the DP-like behaviour.

\begin{figure}[ht]
  {\includegraphics[width=8cm]{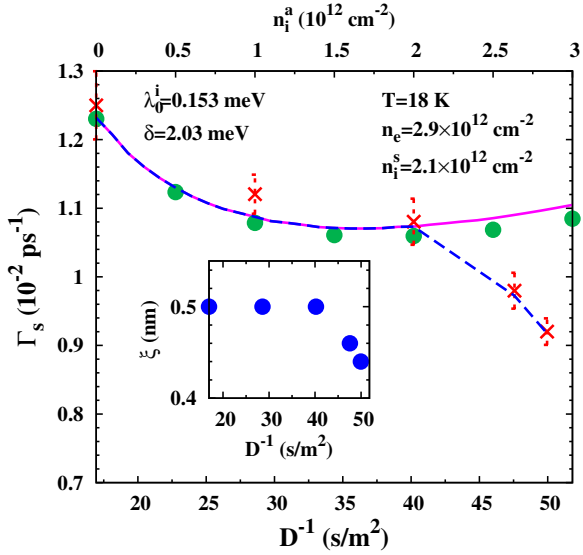}}
  \caption{(Color online) The dependence of in-plane spin relaxation rate on the adatom density (at the top of the frame) or
    the inverse of the charge diffusion coefficient (at the bottom of
    the frame). The temperature is $T=18$~K, the electron density is
    $n_e=2.9\times 10^{12}$~cm$^{-2}$ and the density of impurities in the substrate is $n_i^s=2.1\times
    10^{12}$~cm$^{-2}$. The crosses with error bars are the
    experimental data from Pi {\it et al.}.\cite{Pi} The solid curve
    stands for the fitting data via the single-sided adatom model
    with $\lambda_0^i=0.153$~meV, $\delta=2.03$~meV and $\xi=0.5$~nm. The nearby dots of the
    solid curve are calculated with the spin-flip scattering terms
    removed. The dashed curve is calculated with the same parameters
    as that of the solid curve except that $\xi$ decreases with
    $n_i^a$ when $n_i^a>2\times 10^{12}$~cm$^{-2}$, as shown in the inset.}
  \label{figzw3}
\end{figure}

We further apply the single-sided adatom model to reinvestigate the
experiment of Pi {\it et al.} from Riverside,\cite{Pi} which shows an
obvious DP-like behaviour. At $T=18$~K, with the increasing
density of adatoms (Au atoms) from surface deposition (although Au
atoms also denote electrons to graphene, the electron density is fixed
at 2.9$\times 10^{12}$~cm$^{-2}$ by adjusting the gate voltage\cite{Pi}), the diffusion coefficient decreases while
the spin relaxation time increases, as indicated by the crosses with
error bars in Fig.~\ref{figzw3}. By fitting the experimental
data without adatoms (before Au deposition), we obtain $\lambda_i^0=0.153$~meV and  $n_i^s=2.1\times
10^{12}$~cm$^{-2}$.\cite{zhanggraphene} During Au deposition, a group of parameters,
$\delta=2.03$~meV and $\xi=0.5$~nm,  can reproduce the experimental
data except when the adatom density $n_i^a$ is larger than 2$\times 
10^{12}$~cm$^{-2}$ (the solid curve). By assuming
that $\xi$ decreases with the increase of $n_i^a$ when $n_i^a$ is
large enough (the inset of Fig.~\ref{figzw3}), the recalculation
can cover the experimental data in the region with large
$n_i^a$ (the dashed curve). Similar to Fig.~\ref{figzw2}, the dots nearby the solid curve
are calculated with the spin-flip scattering being removed (with $\xi$
being fixed as 0.5~nm). The small
discrepancy between the solid curve and dots indeed indicates that the
spin-flip scattering due to the RRF is not important.

\subsection{Electron density dependence of spin relaxation}
\begin{figure}[ht]
  {\includegraphics[width=8cm]{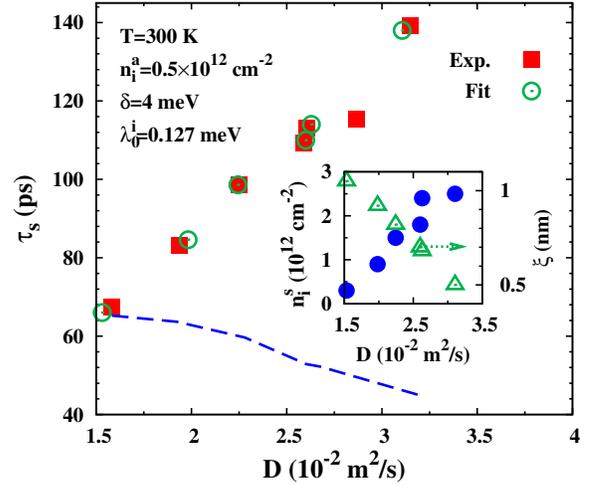}}
  \caption{(Color online) Fit to the experimental data (the dependence
    of spin relaxation time $\tau_s$ on the charge diffusion
    coefficient $D$ with the increase of $n_e$) of J\'ozsa {\it
      et al.}\cite{Jozsa_09} via the single-sided adatom model. The
    squares stand for the experimental data and the open circles are
    from our calculation. The inset shows the dependences of
    $\xi$ (with the scale on the right-hand side of the frame) and
    $n_i^s$ on $D$ when $n_e$ is increased. For comparison, the dashed
    curve is calculated with $\xi$ being fixed at 1.05~nm. $n_i^a=0.5\times 10^{12}$~cm$^{-2}$,
    $\lambda_0^i=0.127$~meV and $\delta=4$~meV.}
  \label{figzw4}
\end{figure}

The electron density dependence of spin relaxation is studied by 
fitting the experiment of J\'ozsa {\it et al.} from
Groningen,\cite{Jozsa_09} which shows an EY-like behaviour. At room
temperature, with the increase of electron density from 0.16 to
2.81$\times 10^{12}$~cm$^{-2}$ (adjusted by the
gate voltage), both the
charge diffusion coefficient and the spin relaxation time increase,
with the latter being proportional to the former (the squares in
Fig.~\ref{figzw4}). It should be noted that this EY-like behaviour can
not be explained by the nearly linear curves shown in
Fig.~\ref{figzw2}, as the linearity there is due to the increase of
adatom density when the electron density is fixed. In fact, according
to Sec.~\ref{analy}, with the increase
of $n_e$ as well as the accompanying increase of
$D$ [and hence the increase of $\tau_p^\ast(k_f)$], both
$\Gamma_{\rm flip}$ and $\Gamma_{\rm DP}$ should increase when the
parameters for the adatom model are fixed [refer to Eqs.~(\ref{flip-limit})
and (\ref{gammadp})]. However, as will be shown in the following, with the
assumption that $\xi$ decreases with increasing $n_e$, both the
single-sided and double-sided adatom models (hence both
the DP mechanism and the mechanism of the
 spin-flip scattering due to the RRF) are able to fit 
the experimental data.  

\begin{figure}[ht]
  {\includegraphics[width=8cm]{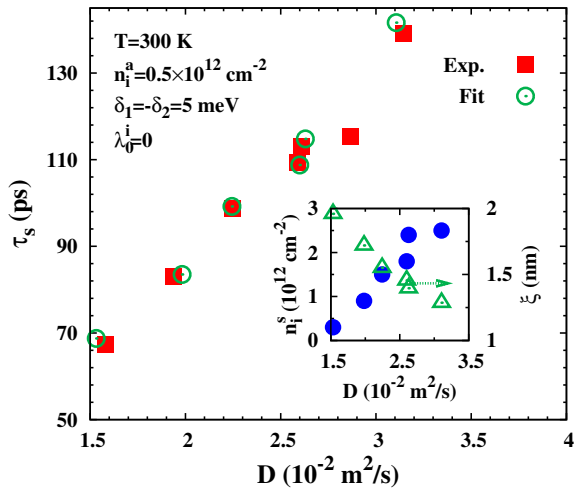}}
  \caption{(Color online) Fit to the experimental data (the dependence
    of spin relaxation time $\tau_s$ on the charge diffusion
    coefficient $D$ with the increase of $n_e$) of J\'ozsa {\it
      et al.}\cite{Jozsa_09} via the double-sided adatom model. The
    squares stand for the experimental data and the open circles are
    from our calculation. The inset shows the dependences of
    $\xi$ (with the scale on the right-hand side of the frame) and
    $n_i^s$ on $D$ when $n_e$ is increased. $n_i^a=0.5\times 10^{12}$~cm$^{-2}$,
    $\lambda_0^i=0$ and $\delta_1=-\delta_2=5$~meV.}
  \label{figzw5}
\end{figure}

In Fig.~\ref{figzw4}, we present the fitting to the experimental data of J\'ozsa {\it et al.}  via the single-sided
adatom model where the DP mechanism is dominant. The calculated spin relaxation time is plotted by the open circles in 
Fig.~\ref{figzw4}. The fitting parameters
are chosen as $\lambda_0^i=0.127$~meV,  $n_i^a=0.5\times 10^{12}$~cm$^{-2}$ and
$\delta=4$~meV here. In order to reproduce 
the electron density dependence of diffusion coefficient, the
impurity density in the substrate $n_i^s$ has to increase with
increasing $n_e$ possibly due to the increased ionization (otherwise if $n_i^s$ is fixed, $D$ will increase
with increasing $n_e$ much more quickly), as shown by the dots in the
inset. Meanwhile, with the increase of $n_e$, $\xi$ should decrease as shown by the triangles in
the inset (the scale is on the right-hand side of the frame) to
account for the increase of $\tau_s$. Otherwise if $\xi$ is fixed, $\tau_s$ will
decrease with increasing $D$ mainly due to the increase of
$\Gamma_{\rm DP}$, as indicated by the dashed curve in the
figure. In Fig.~\ref{figzw5}, we also present a feasible fitting
 by the double-sided adatom model. The squares stand for the experimental data and the open
 circles are from calculation. In our computation, $\lambda_0^i=0$, $n_i^a=0.5\times 10^{12}$~cm$^{-2}$ and
$\delta_1=-\delta_2=5$~meV. The inset shows the dependences of
$\xi$ (open triangles with the scale on the right-hand side of the frame) and
$n_i^s$ (solid circles) on $D$ when $n_e$ is increased. In this fitting, only the
spin-flip scattering due to the RRF plays a role in spin relaxation.

\label{double}

\subsection{Temperature dependence of spin relaxation}
\begin{figure}[ht]
  {\includegraphics[width=8.5cm]{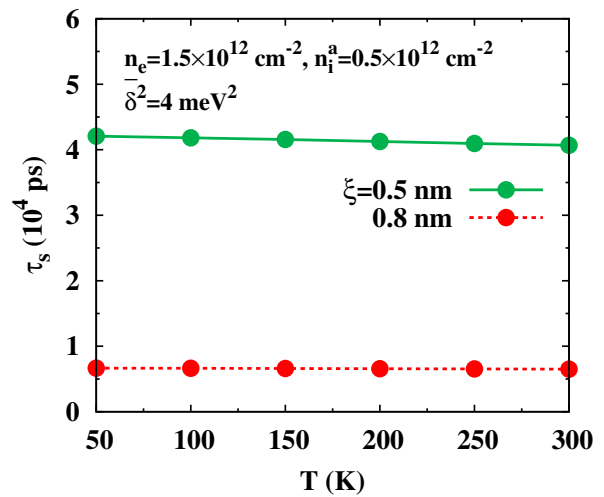}}
  \caption{(Color online) Temperature dependence of spin relaxation
    time due to the spin-flip scattering. The parameters are
    $n_e=1.5\times 10^{12}$~cm$^{-2}$, $n_i^a=0.5\times
    10^{12}$~cm$^{-2}$ and $\bar{\delta^2}=4$~meV$^2$. Solid curve:
    $\xi=0.5$~nm; and dotted curve: $\xi=0.8$~nm.}
  \label{figzw6}
\end{figure}

We investigate the temperature dependence of spin relaxation
in graphene in this section. Although the spin relaxation time
determined by the DP mechanism increases with growing $T$ as
pointed out in Sec.~\ref{analy}, this dependence becomes very week
when the electron-impurity scattering is dominant (which is satisfied in
graphene on SiO$_2$ substrate), as revealed in Fig.~1 of
Ref.~\onlinecite{yzhou} (note the mobility there is even one order
of magnitude larger than the ones in this investigation). 
The spin relaxation time
determined by the spin-flip scattering due to the RRF is also insensitive
to $T$, as shown in Fig.~\ref{figzw6}. Therefore, when other
parameters are fixed, the spin relaxation in graphene is expected to
depend on temperature weakly.

It is quite interesting that a decrease of $\tau_s$ with $T$
is observed by the Riverside group very recently.\cite{han1012.3435} Moreover, when
$T$ is fixed, with the increase of $n_e$
(adjusted by the gate voltage), both $D$ and $\tau_s$ increase,  
similar to the observations by J\'ozsa {\it et al.}.\cite{Jozsa_09} 
The decrease of $\tau_s$ with growing $T$ may be due to the increase
of the correlation length $\xi$ of the RRF with the increase of $T$,
with either the DP mechanism or the spin-flip scattering due to the
RRF being dominant. The linear scaling 
between $\tau_s$ and $D$ with the variation
of electron density also can not determine which mechanism is the
dominant one, as demonstrated in
the previous section. 

\begin{figure}[ht]
  {\includegraphics[width=8.5cm]{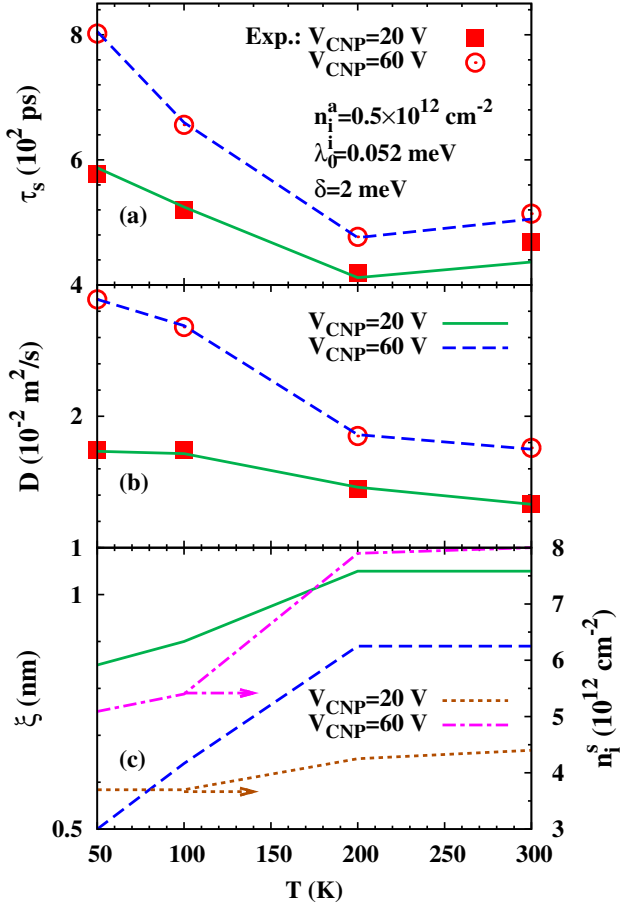}}
  \caption{(Color online) Fit to the experimental data of Han and
    Kawakami\cite{han1012.3435} with the single-sided adatom model. (a) and (b): the temperature
    dependence of $\tau_s$ and $D$. The squares (open circles) are the experimental data with $V_{\rm
      CNP}=20$~V (60~V), and the solid (dashed) curve is from fitting
    with $V_{\rm CNP}=20$~V (60~V). (c): the temperature dependences of
    $\xi$ and $n_i^s$ (on the right-hand side of the frame). The solid
    and dotted curves are for the case with $V_{\rm  CNP}=20$~V, and
    the dashed and chain curves are for the case with $V_{\rm  CNP}=60$~V.}
  \label{figzw7}
\end{figure}

\begin{figure}[ht]
  {\includegraphics[width=8cm]{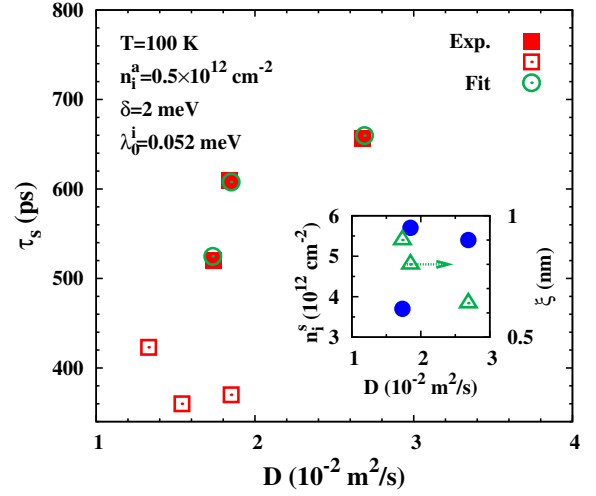}}
  \caption{(Color online) Fit to the experimental data (the dependence
    of spin relaxation time $\tau_s$ on the charge diffusion
    coefficient $D$ with the increase of $n_e$) of Han and
    Kawakami\cite{han1012.3435} via the single-sided adatom model. The
    solid squares stand for the experimental data and the open circles are
    from our calculation. The inset shows the dependences of
    $\xi$ (with the scale on the right-hand side of the frame) and
    $n_i^s$ on $D$ when $n_e$ is increased. The open squares are the
    experimental data for holes or near the charge
    neutrality point, which are not considered in our fitting. $n_i^a=0.5\times 10^{12}$~cm$^{-2}$,
    $\lambda_0^i=0.052$~meV and $\delta=2$~meV.}
  \label{figzw8}
\end{figure}

As a feasible way, we fit the temperature dependence of spin
relaxation based on the single-sided adatom model by assuming $\xi$ to
increase with $T$. One possible fitting 
is shown in Fig.~\ref{figzw7}. Experimentally, when the gate voltage $V_{\rm CNP}=20$~V
(60~V), the electron density $n_e=1.47\times 10^{12}$~cm$^{-2}$
($4.42\times 10^{12}$~cm$^{-2}$).\cite{illu} In Fig.~\ref{figzw7}(a) and
(b), the squares (open circles) are the experimental data of spin
relaxation time and diffusion coefficient with $V_{\rm
  CNP}=20$~V (60~V), respectively, and the solid
(dashed) curves are from our calculation with $V_{\rm   CNP}=20$~V
(60~V). The fitting parameters are $n_i^a=0.5\times
10^{12}$~cm$^{-2}$, $\lambda_0^i=0.052$~meV and $\delta=2$~meV. The
variation of $\xi$ with $T$ is shown in Fig.~\ref{figzw7}(c), where the solid (dashed)
curve is for $V_{\rm   CNP}=20$~V (60~V). The variation of $n_i^s$ with
$T$ is also shown in Fig.~\ref{figzw7}(c) with the scale on the right-hand side of the
frame, where the dotted (chain) curve is for $V_{\rm   CNP}=20$~V
(60~V). In this fitting, the DP mechanism is dominant and the
spin-flip scattering due to the RRF can be neglected. In fact, the calculation with
similar parameters in Fig.~\ref{figzw6} has indicated that the spin
relaxation time caused by the spin-flip scattering due to the RRF is
very long.  With the same parameters, we further fit the
 dependence of spin relaxation time on diffusion coefficient at 100~K
 in Fig.~\ref{figzw8}. In consistence to the fittings in the previous
  section, the correlation length of the RRF is also found to decrease
  with increasing electron density. It is noted that the open squares in the
  figure are the data measured for holes or near the charge
  neutrality point\cite{han1012.3435} and hence
are not considered in our fitting.

\subsection{A nonmonotonic dependence of $\tau_s$ on $D$}
\begin{figure}[ht]
  {\includegraphics[width=8.5cm]{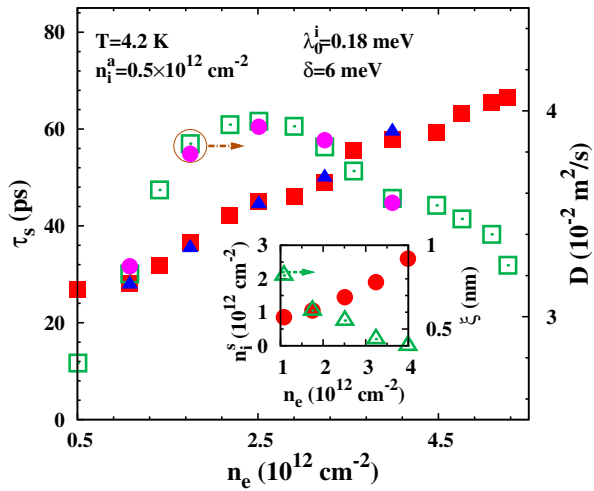}}
  \caption{(Color online) Fit to the experimental data (the dependence
    of spin relaxation time $\tau_s$ and charge diffusion
    coefficient $D$ on electron density $n_e$) of Jo {\it
      et al.}\cite{jo} via the single-sided adatom model. The closed
    squares (solid triangles) are the experimental (fitting) data of
    the spin relaxation time $\tau_s$, and the open squares (solid 
    circles) are the experimental (fitting) data of the diffusion
    coefficient $D$ (with the scale on the right-hand side of the
    frame). The inset shows the density dependence of 
    $\xi$ (with the scale on the right-hand side of the frame) and
    $n_i^s$. $n_i^a=0.5\times 10^{12}$~cm$^{-2}$,
    $\lambda_0^i=0.18$~meV and $\delta=6$~meV.}
  \label{figzw9}
\end{figure}

In the experiments of both Riverside\cite{han1012.3435} and
 Groningen\cite{Jozsa_09} groups, the spin
relaxation time $\tau_s$ is observed to depend on the diffusion
coefficient $D$ monotonically. However, very recently, a nonmonotonic
dependence of $\tau_s$ on $D$ with the increase of carrier
 density at $T=4.2$~K was reported by Jo
{\it et al.}.\cite{jo} Although this phenomenon is reported at 
the hole band, we can treat it  at the electron band by our model
due to the  electron-hole symmetry of band structure in graphene. In
Fig.~\ref{figzw9}, we fit the experimental data by the
single-sided adatom model with $n_i^a=0.5\times 10^{12}$~cm$^{-2}$,
$\lambda_0^i=0.127$~meV and $\delta=4$~meV. The closed
squares (solid triangles) are the experimental (fitting) data of
the spin relaxation time $\tau_s$, and the open squares (solid 
circles) are the experimental (fitting) data of the diffusion
coefficient $D$ (with the scale on the right-hand side of the
frame). The inset shows the density dependence of 
$\xi$ (open triangles with the scale on the right-hand side of the frame) and
$n_i^s$ (solid circles). Due to the slower decrease of $\xi$ and 
 the faster increase
of $n_i^s$ with increasing $n_e$, it is possible for $\tau_s$ to
decrease with $n_e$ when the latter is large enough.

\subsection{Possible factors affecting the correlation length of the RRF}

In our fittings to the experiments, the variation of the correlation
length $\xi$ of the RRF plays an essential role. $\xi$ is found to
decrease with increasing electron density and increase with growing temperature. $\xi$ is also found to decrease with
the increase of adatom density when the latter is high enough. It is 
indeed quite possible
that $\xi$ is affected by these factors. For example, 
the correlation length might be shortened by the screening of carriers,
which is more effective when the carrier density is high. It is also
possible that with growing temperature the adatoms tend to form clusters and enhance the
inhomogeneity.\cite{mccreary} Besides, the puddle size,
which measures the correlation length of the short-range
 random potential in graphene, 
decreases with increasing density of the charged
impurities.\cite{rossi} The similar feature is also expected in
the current study, i.e., $\xi$ decreases with increasing
adatom density when the latter is large enough.

\section{Discussion and conclusion}
\subsection{Discussion on the possible dominant spin relaxation mechanism}
We summarize our numerical fittings to the
experiments and discuss the possible dominant
spin relaxation mechanism. By fitting 
the DP-like behaviour with the increase of adatom
density observed by the Riverside group,\cite{Pi} we find that the
DP mechanism is the dominant one, only with which can the experimental
phenomenon be understood. It is noted that the correlation length
$\xi$ of the RRF is supposed to be constant at the low
density regime of adatoms, and when $\xi$ is fixed $\Gamma_{\rm flip}$
always increases with the adatom density. Therefore, other kinds of
attempts with the spin-flip scattering due to the RRF being dominant
fail to reproduce the experimental phenomenon and can be
ruled out. However, the  EY-like
behaviour with the increase of electron density observed by the
Groningen group\cite{Jozsa_09} can be
fitted by our model with either the DP mechanism or the spin-flip scattering 
due to the RRF being dominant when the decrease of $\xi$ with 
increasing electron density is considered. Nevertheless, the fact
that Riverside group has also observed the similar EY-like behaviour
in their samples very recently,\cite{han1012.3435}
in combination with the observation of the adatom density dependence
of the spin relaxation,\cite{Pi}  suggests that the DP mechanism is
dominant, but exhibiting EY-like properties with the increase of
electron density. The similar experimental
phenomenon on the electron density dependence of the 
spin relaxation from the Groningen\cite{Jozsa_09} and 
Riverside\cite{han1012.3435} groups further
suggests that the DP mechanism also dominates the spin relaxation in the
experiment of Groningen group.\cite{Jozsa_09} Consequently, with the
DP mechanism being dominant in graphene, the RRF leads to spin 
relaxation which exhibits either DP- or EY-like properties in the
experiments. 

\subsection{Conclusion}
In conclusion, we have studied electron spin relaxation in graphene with
random Rashba field by means of the kinetic spin Bloch equations,
aiming to understand the main spin relaxation mechanism in
  graphene of the current experiments. Different from the previously studied case
by Zhou and Wu where no adatoms are considered and the mobility is relatively
 high,\cite{yzhou,novoselov666} the electron mobility investigated in the
present work is at least one order of
 magnitude smaller due to the extrinsic factors caused by the ferromagnetic 
electrodes used in the spin relaxation
 measurements,\cite{Tombros_08,han222109,Popinciuc,Jozsa_09,han1012.3435,jo} 
e.g., the  adatoms.  We set up a
random Rashba model to incorporate the contribution from both
the adatoms and substrate. In this model, the charged adatoms on one hand
enhance the Rashba spin-orbit coupling locally and on the other hand serve as
Coulomb potential scatterers. 

Based on the random Rashba model, the analytical study on spin
relaxation in graphene is performed. The average of the random Rashba field leads to spin
relaxation limited by the D'yakonov-Perel' mechanism, which is absent in the
work by Dugaev {\it et al.},\cite{dugaev085306} while the randomness of the
random Rashba field causes spin relaxation by spin-flip scattering, serving as an
Elliott-Yafet--like mechanism. With the increase of adatom density, the spin relaxation caused by the
spin-flip scattering due to the random Rashba field always shows an
Elliott-Yafet--like behaviour, whereas the D'yakonov-Perel' mechanism can exhibit
either Elliott-Yafet-- or D'yakonov-Perel'--like one. When all the other parameters are
fixed, with the increase of electron density the spin
relaxation rates due to both mechanisms increase;
Nevertheless, the spin relaxation rate 
determined by the spin-flip scattering due to the random Rashba field
is insensitive to the temperature whereas that determined by the D'yakonov-Perel'
mechanism becomes insensitive to the temperature when the electron-impurity
scattering is dominant. However, both mechanisms are
 sensitive to the correlation length of the random Rashba field, which may be affected
 by the environmental factors such as electron density and temperature. 

We further carry out numerical calculations and fit the experiments of
Riverside\cite{Pi,han1012.3435} and Groningen\cite{Jozsa_09}
groups, which show either D'yakonov-Perel'-- or Elliott-Yafet--like property. 
By fitting and comparing these experiments, we
suggest that the D'yakonov-Perel' mechanism dominates the spin relaxation in
graphene. With the D'yakonov-Perel' mechanism being dominant, the random Rashba field leads to spin 
relaxation which exhibits either D'yakonov-Perel'-- or Elliott-Yafet--like properties.  
Besides, a latest reported nonmonotonic dependence of
$\tau_s$ on $D$ by Jo {\it et al.}\cite{jo} is also fitted by our
 model with the D'yakonov-Perel' mechanism being dominant. 

\begin{acknowledgments}

This work was supported by the
National Basic Research Program of China under Grant No.\,2012CB922002
and the Natural Science Foundation of China
under Grant No.\ 10725417.

\end{acknowledgments}

\begin{appendix}
\section{Spin-flip scattering terms}
\label{app-a}
The spin-flip scattering terms are\cite{glazov2157}
\begin{eqnarray}\nonumber
\partial_t\rho_{\mu{\bf k}}(t)|_{\rm scat}^f&=&-\frac{\pi}{\hbar}\sum_{{\bf k}^\prime\neq {\bf k}} |\lambda_{\bf
  k-k^\prime}|^2\delta(\varepsilon_{\mu\bf k}-\varepsilon_{\mu\bf
  k^\prime})\\ \nonumber &&\mbox{}\times [\rho_{ \mu{\bf k}}(t)V_{\bf
  kk^\prime}V_{\bf k^\prime k}+V_{\bf
  kk^\prime}V_{\bf k^\prime k}\rho_{ \mu{\bf k}}(t)\\\nonumber&&\mbox{}-2V_{\bf kk^\prime}\rho_{\mu{\bf
  k^\prime}}(t)V_{\bf k^\prime k}]\\\nonumber
&=&-\frac{2\pi}{\hbar}\sum_{{\bf k}^\prime} |{\tilde \lambda}_{\bf
  k-k^\prime}|^2\delta(\varepsilon_{\mu\bf k}-\varepsilon_{\mu\bf
  k^\prime})\\  &&\mbox{}\times [\rho_{ \mu{\bf k}}(t)-V_{\bf kk^\prime}\rho_{\mu{\bf
  k^\prime}}(t)V_{\bf k^\prime k}],
\label{oflip}
\end{eqnarray}
where
\begin{eqnarray}
{\tilde \lambda}_{\bf q}&=&\int [\lambda({\bf r})-
\lambda_0]e^{-i{\bf q}\cdot{\bf r}}d{\bf r}, 
\end{eqnarray}
satisfying ${\tilde \lambda}_{{\bf q}=0}=0$ and ${\tilde
  \lambda}_{{\bf q}\neq 0}=\lambda_{{\bf q}}$. When the mean free path $l$ is 
much larger than the correlation
length $\xi$ of the fluctuating Rashba field, $|{\tilde
  \lambda}_{\bf q}|^2$ can be approximated by its statistical
average as follows,
\begin{eqnarray}\nonumber
|{\tilde \lambda}_{\bf q}|^2
&\approx&\int \int d{\bf
  r}d{\bf r^\prime}\langle[\lambda({\bf r})-\lambda_0][\lambda({\bf
  r^\prime})-\lambda_0]\rangle\\ \nonumber &&\mbox{}\times e^{-i{\bf
    q}\cdot({\bf r}-{\bf r^\prime})}\\
&=& \int \int d{\bf
  r}d{\bf r^\prime}C({\bf r}-{\bf r^\prime})e^{-i{\bf
    q}\cdot({\bf r}-{\bf r^\prime})}=C_{\bf q}.
\label{cq}
\end{eqnarray}
Eqs.~(\ref{oflip}) and (\ref{cq}) lead to Eq.~(\ref{flip}).

\section{Analytical study of spin relaxation in graphene}
\label{app-b}
We present the analytical study of spin relaxation in
graphene in detail. By expanding ${\bf S}_{\bf k}(t)$ as ${\bf S}_{\bf  k}(t)=\sum_{l}{\bf
  S}_{k}^l(t)e^{il\theta_{\bf k}}$, one obtains from
Eq.~(\ref{analy-ksbe}) the following equations,

\begin{eqnarray}\nonumber
\partial_t{S}^l_{kx}(t)=&&\frac{\lambda_0}{\hbar}\sum_{l_0=\pm
  1}S_{kz}^{l+l_0}(t)-(\frac{1}{\tau_{k}^l}+\frac{1}{\tau_{ks}^0})S_{kx}^l(t)\\
&&\mbox{}-\sum_{l_0=\pm
1}\frac{S_{kx}^{l+2l_0}(t)+i^{l_0}S_{ky}^{l+2l_0}(t)}{2\tau_{ks}^{l+l_0}},\label{sx}\\\nonumber
\partial_t{S}^l_{ky}(t)=&&\frac{\lambda_0}{\hbar}\sum_{l_0=\pm
  1}i^{l_0}S_{kz}^{l+l_0}(t)-(\frac{1}{\tau_{k}^l}+\frac{1}{\tau_{ks}^0})S_{ky}^l(t)\\
&&\mbox{}-\sum_{l_0=\pm
  1}\frac{i^{l_0}S_{kx}^{l+2l_0}(t)-S_{ky}^{l+2l_0}(t)}{2\tau_{ks}^{l+l_0}},\label{sy}\\\nonumber
\partial_t{S}^l_{kz}(t)=&&-\frac{\lambda_0}{\hbar}\sum_{l_0=\pm
  1}[S_{kx}^{l+l_0}(t)+i^{l_0}S_{ky}^{l+l_0}(t)]\\ &&
\mbox{}-(\frac{1}{\tau_{k}^l}+\frac{2}{\tau_{ks}^0})S_{kz}^l(t),
\label{sz}
\end{eqnarray}
in which
\begin{equation}
\frac{1}{\tau^l_k}=\frac{k}{4\pi\hbar^2v_f}\int_0^{2\pi}d\theta
|U_{\bf q}|^2(1+\cos\theta)(1-\cos
l\theta)
\label{taukl}
\end{equation}
and 
\begin{equation}
\frac{1}{\tau^l_{ks}}=\frac{k}{2\pi\hbar^2v_f}\int_0^{2\pi}d\theta C_{\bf q}\cos l\theta.
\label{tauksl}
\end{equation}
Here $|U_{\bf q}|^2$ and $C_{\bf q}$ depend only on $|{\bf q}|=2k\sin\frac{\theta}{2}$. It is
noted that $\frac{1}{\tau^l_k}=\frac{1}{\tau^{-l}_k}$ and 
$\frac{1}{\tau^l_{ks}}=\frac{1}{\tau^{-l}_{ks}}$. It is also noted that $\frac{1}{\tau^0_k}=0$
and $\tau_k^1$ is in fact the momentum relaxation time $\tau_p(k)$ limited by the
electron-impurity scattering. 

By retaining the lowest three orders of ${\bf S}_{k}^l(t)$ (i.e., 
terms with $l=0$, $\pm 1$) in Eqs.~(\ref{sx})-(\ref{sz}), one obtains
\begin{eqnarray}
\left[\partial_t-\left(
  \begin{array}{ccc}
    {\bf F} & {\bf P} & {\bf Q}\\
    -{\bf P}^\dagger & {\bf G} & {\bf P}\\
    {\bf Q}^\dagger & -{\bf P}^\dagger & {\bf F}
    \end{array}
\right)\right]\left(
  \begin{array}{c}
    {\bf S}_k^1\\
    {\bf S}_k^0\\
    {\bf S}_k^{-1}
    \end{array}
\right)=0,
\label{sequation}
\end{eqnarray}
where 
\begin{eqnarray}
&&\hspace{-0.5 cm}{\bf G}=-\frac{1}{\tau_{ks}^0}\left(\begin{array}{ccc}
    1 & 0 & 0 \\
    0 & 1 & 0 \\
    0 & 0 & 2
\end{array}
\right),\\
&&\hspace{-0.5 cm} {\bf F}={\bf G}-\frac{1}{\tau_k^1},\\
&&\hspace{-0.5 cm}{\bf P}=\frac{\lambda_0}{\hbar}\left(\begin{array}{ccc}
    0 & 0 & 1 \\
    0 & 0 & -i \\
    -1 & i  &  0
\end{array}
\right),\\
&&\hspace{-0.5 cm}{\bf Q}=\frac{1}{2\tau_{ks}^0}\left(\begin{array}{ccc}
    -1 & i & 0 \\
     i & 1 & 0 \\
    0 & 0  &  0
\end{array}
\right).
\end{eqnarray}
As the spin flipping rate $1/\tau_{ks}^0$ is much smaller than the
momentum relaxation rate $1/\tau_{k}^1$ (in graphene $1/\tau_k^1$ 
is usually of the order of 10~ps$^{-1}$; even if $1/\tau_{ks}^0$ reaches the experimental value $\sim$ 0.01~ps$^{-1}$,
$\tau_{k}^1/\tau_{ks}^0$ is still as
small as 10$^{-3}$) and ${\bf
  S}_k^{\pm 1}$ are smaller terms compared to ${\bf S}_k^0$ in
the strong scattering limit, we
approximate ${\bf F}$ and ${\bf Q}$ as ${\bf F}\approx
-\frac{1}{\tau_k^1}$ and
${\bf Q}\approx 0$.  With initial conditions, e.g., ${\bf S}_k^l(0)=\delta_{l0}(0,0,S_{kz}^0(0))^T$ ($l=0,\pm 1$), Eq.~(\ref{sequation}) can be
solved as [those of ${\bf  S}_k^{\pm
  1}(t)$ are away from our interest and are not shown here]
\begin{eqnarray}\nonumber
{\bf S}_{k}^0(t)&=&\frac{1}{2}\Big[\Big(1+\frac{1}{\sqrt{1-c_z^2}}\Big)e^{-\Gamma_z^+t}
+\Big(1-\frac{1}{\sqrt{1-c_z^2}}\Big)\\\mbox{}&&\times e^{-\Gamma_z^-t}\Big](0,0,S_{kz}^0(0))^T,
\end{eqnarray}
where 
\begin{equation}
c_z=\frac{4\lambda_0}{\hbar}\frac{1}{1/\tau_k^1-1/\tau_{ks}^0}
\end{equation}
and 
\begin{equation}
\Gamma_z^\pm=\frac{1}{\tau_{ks}^0}+\frac{1}{2\tau_k^1}\pm(\frac{1}{\tau_{ks}^0}-\frac{1}{2\tau_k^1})\sqrt{1-c_z^2}.
\end{equation} 
In the strong scattering limit with $c_z\ll 1$, $S_{kz}^0(t)\approx
S_{kz}^0(0)e^{-\Gamma_z^+t}$ where $\Gamma_z^+\approx
2/\tau_{ks}^0+4\lambda_0^2\tau_k^1/\hbar^2$. Consequently, one obtains
\begin{eqnarray}
{\bf S}_k^0(t)=e^{-\Gamma_zt}(0,0,S_{kz}^0(0))^T
\end{eqnarray}
with 
\begin{eqnarray}
 \Gamma_z=2/\tau_{ks}^0+4\lambda_0^2\tau_k^1/\hbar^2.
\end{eqnarray}  
Similarly, with ${\bf S}_k^l(0)=\delta_{l0}(S_{kx}^0(0),0,0)^T$ or ${\bf
  S}_k^l(0)=\delta_{l0}(0,S_{ky}^0(0),0)^T$, ${\bf S}^0_k(t)$ is
solved to be 
\begin{eqnarray}
{\bf S}_k^0(t)=e^{-\Gamma_xt}(S_{kx}^0(0),0,0)^T, 
\end{eqnarray}
and 
\begin{eqnarray}
{\bf S}_k^0(t)=e^{-\Gamma_yt}(0,S_{ky}^0(0),0)^T,
\end{eqnarray}
respectively, with $\Gamma_x=\Gamma_y=\Gamma_z/2$.

\end{appendix}

\end{document}